\documentclass[runningheads]{llncs}

\usepackage[T1]{fontenc}
\usepackage{graphicx}
\usepackage{array}
\usepackage{rotating} 
\usepackage{booktabs}
\usepackage{diagbox}
\usepackage{multirow}
\usepackage{marvosym}
\usepackage{xcolor}
\usepackage{arydshln}
\usepackage{algorithm}
\usepackage{algpseudocode}
\usepackage{amsmath}
\usepackage{makecell}
\usepackage{nicematrix}
\usepackage[colorlinks=true,
            linkcolor=red,    % Internal links (sections, figures)
            citecolor=blue,   % Citation links
            urlcolor=blue,    % URLs
            pdfborder={0 0 0}]{hyperref} % Remove all link borders

\usepackage{orcidlink}

\begin{document}

\title{Leakage-abuse Attack Against Substring-SSE with Partially Known Dataset}

\author{
    Xijie Ba$^{1}$\orcidlink{0009-0000-5105-2124} \and
    Qin Liu$^{1,\textsuperscript{\Letter}}$\orcidlink{0000-0002-8979-5094} \and
    Xiaohong Li$^{2,\textsuperscript{\Letter}}$\orcidlink{0000-0003-1169-6974} \and
    Jianting Ning$^{1}$\orcidlink{0000-0001-7165-398X}
}
\institute{
$^{1}$School of Cyber Science and Engineering, Wuhan University, Wuhan, China\\
\email{\{baxijie,qinliu,jtning\}@whu.edu.cn}\\
$^{2}$School of Computer Science, Wuhan University, Wuhan, China\\
\email{leexh@whu.edu.cn}
}

% \author{}  % 显式留空
% \institute{} % 如果需要也隐藏institute

\maketitle              % typeset the header of the contribution
% \footnotetext{The work was supported by the National Natural Science Foundation of China (No. 62272348).}

\begin{abstract}
Substring-searchable symmetric encryption (substring-SSE) has become increasingly critical for privacy-preserving applications in cloud systems. 
However, existing schemes remain vulnerable to information leakage during search operations, particularly when adversaries possess partial knowledge of the target dataset. 
%While leakage-abuse attacks have been extensively studied for conventional SSE schemes, no prior work has systematically investigated such attacks against substring-SSE in realistic partially-known scenarios.
Although leakage-abuse attacks have been widely studied for traditional SSE, their applicability to substring-SSE under partially known data assumptions remains unexplored.

In this paper, we present the first leakage-abuse attack on substring-SSE under partially-known dataset conditions. 
We develop a novel matrix-based correlation technique that extends and optimizes the LEAP framework for substring-SSE, enabling efficient recovery of plaintext data from encrypted suffix tree structures. 
Unlike existing approaches that rely on independent auxiliary datasets, our method directly exploits known data fragments to establish high-confidence mappings between ciphertext tokens and plaintext substrings through iterative matrix transformations.

Comprehensive experiments on real-world datasets demonstrate the effectiveness of the attack, with recovery rates reaching 98.32\% for substrings given 50\% auxiliary knowledge. 
Even with only 10\% prior knowledge, the attack achieves 74.42\% substring recovery while maintaining strong scalability across datasets of varying sizes. 
The result reveals significant privacy risks in current substring-SSE designs and highlights the urgent need for leakage-resilient constructions.

\end{abstract}
\section{Introduction}
%\subsection{A Subsection Sample}
The rapid advancement of cloud computing\cite{bello2021cloud} and big data analytics\cite{himeur2023ai} has revealed functional limitations in traditional encryption methods for data storage (e.g., AES\cite{mendonca2018data}, SSL/TLS\cite{satapathy2016comprehensive}), as their design inherently restricts efficient search operations.
%In response to this challenge, Searchable Encryption (SE)\cite{andola2022searchable} emerges as a foundational primitive to enable secure queries over encrypted data while preserving confidentiality.
In response to this challenge, searchable symmetric encryption (SSE)\cite{andola2022searchable,curtmola2006searchable,gui2023rethinking,li2023survey} enables data owners to outsource storage while preserving privacy, allowing clients to store and distribute large volumes of symmetrically encrypted data at low cost, while maintaining controlled retrieval access to the encrypted data through secure search protocols.

Substring-searchable symmetric encryption (substring-SSE) enables substring search over encrypted database. 
Its primary applications include secure email systems and encrypted DNA sequence searches.
Compared to traditional SSE, substring-SSE overcomes the limitation of fixed keywords by supporting arbitrary substring queries\cite{10.1145/3064005}. 

Since the first substring-SSE scheme\cite{chase2015substring} was proposed by Chase and Shen, the direction has witnessed multiple substring-SSE schemes emerging\cite{faber2015rich,hahn2018practical,hoang2023novel,leontiadis2018storage}.
%The direction has witnessed multiple novel substring-SSE schemes emerging,such as\cite{boldyreva2014efficient，chase2015substring，faber2015rich,hahn2018practical,hoang2023novel,leontiadis2018storage，hoang2023novel}.
%实验性能可以借助越南的方案，子串搜索方案对比（对称&非对称）
Its core functionality enables efficient retrieval of all matched positions for arbitrary target substrings through specific encrypted data structure\cite{chase2015substring,leontiadis2018storage} or transfer substring queries to range queries or conjunctive keyword queries\cite{faber2015rich,hahn2018practical,hoang2023novel}. 
However, substring-SSE schemes face significant challenges in leakage suppression when enabling efficient privacy-preserving retrieval. 
To the best of our knowledge, there is no prior work that provides leakage suppression technique for substring-SSE schemes, which has led to the emergence of leakage-abuse attacks. 
These attacks exploit scheme-induced leakage to compromise either data privacy or query privacy. 
For instance, two substrings may partially overlap (e.g., "abc" and "bcd" share "bc"), allowing the server to infer character relationships from leaked matching positions and launch fine-grained data reconstruction attacks. 
Although numerous leakage-abuse attacks have been proposed for various SSE schemes (e.g., \cite{lambregts2022val,markatou2021reconstructing,ning2021leap,9165915,8443434,xu2021interpreting,xu2023leakage}), 
such attacks are rarely explored for substring-SSE due to its prohibitive computational/storage overhead and the complexity of substring segmentation.
Table \ref{tab:substring_sse} gives an overview of existing substring-SSE schemes.

\begin{table}[h]
\centering
\caption{Comparative analysis of substring-SSE schemes: extensions from traditional SSE, leakage profiles, corresponding known attacks, and underlying data structures.}
\label{tab:substring_sse}
\begin{tabular}{l>{\raggedright\arraybackslash}p{3.2cm}>{\raggedright\arraybackslash}p{3cm}cc}
\toprule
\textbf{Schemes} & \textbf{Extension} & \textbf{Leakage} & \textbf{Attacks\ } & \textbf{\ Structure} \\ 
\midrule
ESORICS~\cite{faber2015rich} 
& substring, wildcard, phrase, boolean search 
& co-occurrence / access pattern 
& --- 
& binary tree \\

SIGMOD~\cite{hahn2018practical} 
& DB-compatible substring search 
& prefix / index intersection pattern 
& --- 
& k-gram \\

DIQ-SSE~\cite{hoang2023novel} 
& low-FP substring search 
& prefix / volume pattern 
& --- 
& suffix tree \\

S3E~\cite{leontiadis2018storage} 
& dynamic substring search 
& access/volume pattern 
& --- 
& suffix array \\
\bottomrule
\end{tabular}
\end{table}

In 2024, Zichen Gui et al. \cite{gui2024query} proposed the first query reconstruction attack against substring-SSE schemes, specifically targeting the Chase-Shen scheme. 
Their key innovation lies in leveraging an \textbf{independent auxiliary dataset} that shares distributional properties with the target dataset. 
By constructing suffix trees from both datasets and analyzing leakage patterns, they establish statistical correlations between ciphertext queries and plaintext substrings. 
With 50\% auxiliary data, the attack employs simulated annealing to optimize the matching process, achieving up to 72\% string recovery rate on genomic data and 49-66\% on English texts. 
However, its performance depends heavily on the Independent and Identically Distributed Assumption (\textbf{IID assumption}) between auxiliary and target data, which requires careful tuning of parameters, such as $\epsilon$ (candidate set size) and $t$ (trimming threshold).  
Its limitations in distributional dependency motivate our enhanced approach.

% In 2024, Zichen Gui et al. proposed the first attack against substring-SSE schemes, demonstrating a query reconstruction attack targeting the Chase-Shen scheme \cite{gui2024query}.
% In this paper, the authors implement string segmentation and representation based on a suffix tree structure. 
% They partition the leakage generated by the CS scheme into two parts: one assigned to an auxiliary database and the other to the target database. 
% By observing query inputs and outputs, they infer similarity or equivalence information between different queries. 
% Finally, they employ simulated annealing to obtain the ultimate matching results.
% However, the query recovery rate of this scheme remains around 55\% as the number of strings increases, which is not particularly ideal.
% To address this, our scheme further explores approaches to improve the query recovery rate for substring-SSE schemes.

\subsection{Our Contribution} 
In this paper, we first introduce a generic architecture for substring-SSE, outlining the communication flow, encryption principles, and search mechanisms. We then analyze leakage generation in existing substring-SSE schemes.

We propose a leakage-abuse attack that utilizes partially known datasets. Our approach adopts the string segmentation and suffix tree structure introduced in Gui et al. \cite{gui2024query} for database initialization, and extends the LEAP attack methodology \cite{ning2021leap} to substring-SSE schemes. By representing database entries and their relationships in matrix form and identifying mappings through row and column transformations, our method enables efficient and effective data recovery.

Finally, we conduct experiments on the Enron dataset, demonstrating that our method achieves 97.87\% alphabet recovery, 98.32\% string recovery, and 94.22\% initial path recovery with 50\% auxiliary knowledge, and achieve 100\% recovery with 60\% knowledge. Notably, even with only 10\% prior knowledge, the attack attains 65.96\% alphabet recovery and 74.42\% string recovery. Robustness evaluation further shows that recovery rates drop by less than 5\% as the dataset scale increases from 1,000 to 30,000 strings, confirming the practical effectiveness and scalability of the proposed attack.

This paper is organized as follows. 
Section~\ref{sec:background} provides the necessary cryptographic preliminaries and related work. 
The universal architecture for substring-SSE schemes is established in Section~\ref{sec:system_model}. 
Section~\ref{sec:attack} details our novel attack methodology against substring-SSE implementations. 
Experimental evaluation and security analysis are presented in Section~\ref{sec:evaluation}. 
We conclude with discussions and future work in Section~\ref{sec:conclusion}. 
The algorithmic implementation of the proposed attack is provided in the Appendix~\ref{sec:appendix}.

\section{Preliminaries}\label{sec:background}
\subsection{SSE and Substring-SSE}

Searchable Symmetric Encryption (SSE) supports keyword search, where a client encrypts a set of documents and can later query them using keywords to retrieve documents containing the specified keyword \cite{song2000practical,curtmola2006searchable,amorim2023leveraging}. 
%The core idea involves encrypting the association between keywords and documents while enabling privacy-preserving searches through tokens. 
However, SSE's keyword search cannot directly support substring search due to the quadratic growth of the substring combination space $(O(n^2))$ \cite{chase2015substring}. 
Treating every possible substring as an independent keyword would lead to prohibitive storage and computational overhead. 

To address this, substring-SSE is proposed by Chase and Shen \cite{chase2015substring} to enable substring search. 
Specifically, given an encrypted string $es$, substring-SSE allows to perform substring queries and returns all occurrence positions of the strings including $es$. 
By leveraging the suffix tree data structure, substring-SSE achieves substring search efficiency comparable to that in plaintext scenarios. 
A substring-SSE scheme consists of the following polynomial-time algorithms.

\begin{itemize}
    \item $k \leftarrow \mathbf{ {Gen}}(1^\lambda)$ : Data owner inputs security parameter $\lambda$ and outputs secret key $k$, which is distributed to data users.
    \item $SC \leftarrow  \mathbf{{Enc}}(k,f)$: Data owner inputs $k$ and files $f \in \mathcal{F}^*$, outputs searchable ciphertext $SC$, which is uploaded to the server. 
    \item $F(es) \leftarrow \mathbf{{Search}}(k,s)$: Client computes $es$ using $k$ and string $s \in \mathcal{S}^*$.
        Then the server retrieves and outputs $F(es)$, i.e., the set of substring indices in $s$.
\end{itemize}
The scheme satisfies correctness: for all $\lambda$, $s$, and $f$, if $k \leftarrow \mathbf{ {Gen}}(1^\lambda)$ and $SC \leftarrow  \mathbf{{Enc}}(k,f)$, then $\mathbf{{Search}}(k,s)$ outputs ${F}(es)$ with overwhelming probability.

% \subsection{CS Scheme}
% \subsubsection{Technical Overview}
% \subsubsection{Leakage of CS Scheme}
% \subsubsection{Query Reconstruction Attack}
\subsection{LEAP Scheme}
At CCS 2021, Ning et al.\cite{ning2021leap} proposed a leakage-abuse attack against traditional SSE that operates with partial knowledge of the dataset (\textbf{LEAP}). 
By leveraging partial knowledge in efficiently deployable, efficiently searchable encryption (EDESE) schemes characterized by Cash et al.\cite{cash2015leakage}, LEAP employs a recursive matrix row/column mapping technique. 
This approach achieves zero false positives of query-to-keyword mappings for the first time.
The scheme provides us with novel insights for data processing with partially known dataset.

% \subsubsection{Hamming Weight}

\subsubsection{Goal}
The LEAP attack is conducted from EDESE schemes. 
Its objective is to recover the mapping of encrypted documents and documents $\mathbf{(ed, d)}$, and the mapping of query token and keyword $\mathbf{(q, w)}$ using leakage and partial knowledge of the target scheme.

\subsubsection{Technical Overview}

LEAP assumes the adversary has access to partial plaintext documents $F^{'}=\{d_{1},...,d_{n}\}$ and keywords $W^{'}=\{w_{1},...,w_{m}\}$.
Leveraging $F^{'}$ and $W^{'}$, the attacker constructs mapping and occurrence matrices as data preparation. 
The attack scheme subsequently executes following five steps.
\begin{itemize}
    \item \textbf{Unique column-sum mapping}. 
    Given that the row sums of the extended $(d,w)$-matrix are identical to those of the $(ed,q)$-matrix, an attacker can establish unique column-sum correspondences between these matrices, thereby recovering partial $(ed,d)$ mappings.
    \item \textbf{Occurrence matrix mapping}. 
    Leveraging the partial $(ed,d)$ mappings obtained in Step~1, the attacker can exploit the relationship between the $n\times n$ $ed$-occurrence matrix $M$ and the $n'\times n'$ $d$-occurrence matrix $M'$ to deduce additional $(ed,d)$ mappings.
    \item \textbf{Unique row mapping}. 
    Since the column-rearranged $(ed,q)$ matrix and the extended column-rearranged $(d,w)$-matrix have identical column sums, the attacker can leverage existing $(ed,d)$ mappings to establish unique row correspondences, thereby recovering partial $(q,w)$ mappings.
    \item \textbf{Unique column mapping}. 
    When unique columns of the row-rearranged $(ed,q)$-matrix map those of the extended row-rearranged $(d,w)$-matrix with, the attacker can identify unique column correspondences to acquire further $(ed,d)$ mappings.
    \item \textbf{Iterative recovery}. 
    The attacker can iteratively recover $(q,w)$ mappings until convergence (i.e., no additional $(q,w)$ mappings can be discovered).
\end{itemize}

% \subsubsection{Motivation}
% Our scheme is motivated by the LEAP attack,

The LEAP scheme provides a novel direction for our research. By adapting and optimizing this matrix mapping methodology for attacks against substring-SSE schemes, we can significantly improve the recovery rates for token characters, initial paths, and strings.

\section{Architecture for substring-SSE}\label{sec:system_model}

\subsection{System Model}
%\subsubsection{Overview}
The following gives the system model of the substring-SSE. 
The entities include Data Owner (DO), Data User (DU) and  Cloud Server (CS). 
DO is responsible for key management and retains exclusive write access to the encrypted database. 
DUs are permitted to submit substring queries to the CS. 
The CS stores the searchable ciphertexts generated and uploaded by the DO and processes query requests, returning the corresponding encrypted results to the DUs.

\begin{figure}
\includegraphics[width=\textwidth]{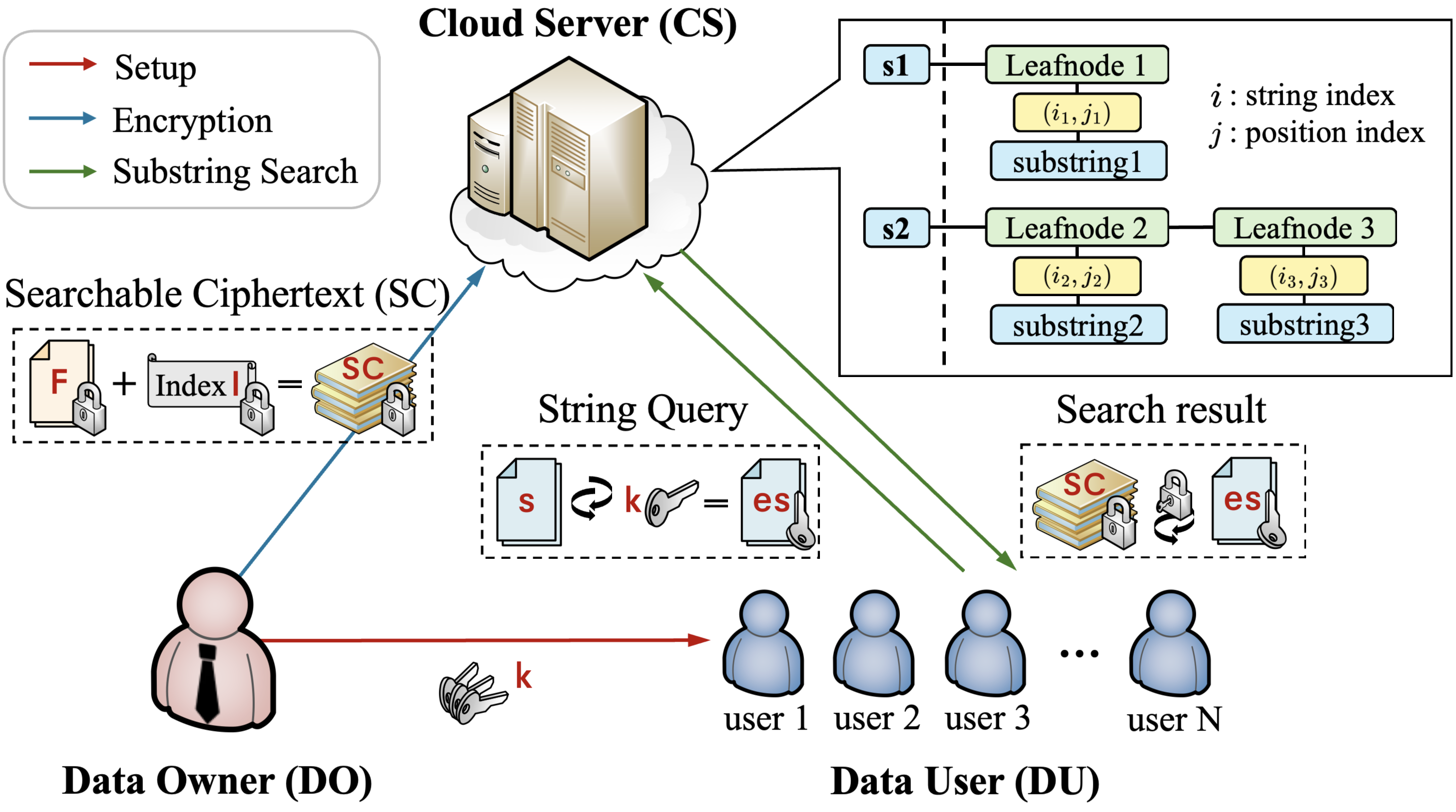}
\caption{The system model of substring searchable symmetric encryption.} \label{fig1}
\end{figure}

The process consists of three primary phases: \textbf{Setup}, \textbf{Encryption}, and \textbf{Substring Search}.
During the \textbf{Setup} phase, DO securely distributes secret keys $k$ to authorized DUs. %while maintaining a local copy, which is used for both data encryption and decryption. 
During the \textbf{Encryption} phase, DO constructs searchable ciphertexts (SC) through processing of data documents and suffix tree index file, both are encrypted.
Each document contains multiple strings, and the suffix tree index file facilitates substring retrieval. 
Finally, DO uploads the generated SC to CS. 
During the \textbf{Substring Search} phase, DU submits an encrypted string query $es$ to CS. 
Then CS retrieves the suffix tree using both the $es$ and SC to locate the corresponding leaf nodes. 
Subsequently, CS retrieves and returns the search results to DU according to the encrypted suffix tree file, including substrings positions. 
%Notably, under honest assumptions, CS cannot access the plaintext data of either the SC or $es$ during the entire retrieval process.

\subsubsection{Suffix Tree}
% \subsection{Suffix Trees}
Suffix trees serve as a fundamental data structure in substring-SSE \cite{chase2015substring,leontiadis2018storage}, offering efficient solutions for complex query operations \cite{gao2005psist,guo2024gridse,strizhov2015substring}. 
In simpler terms, the suffix tree used in substring-SSE can be regarded as analogous to the encrypted multi-maps (EMM) in traditional SSE schemes, both serve as encrypted dictionaries. 
The difference lies in their mapping: EMM maps document keywords to corresponding document identifiers, the suffix tree in substring-SSE maps a string to the indices associated with its substrings shown in Figure~\ref{fig1}.

\begin{figure}
\centering
\includegraphics[width=0.7\textwidth]{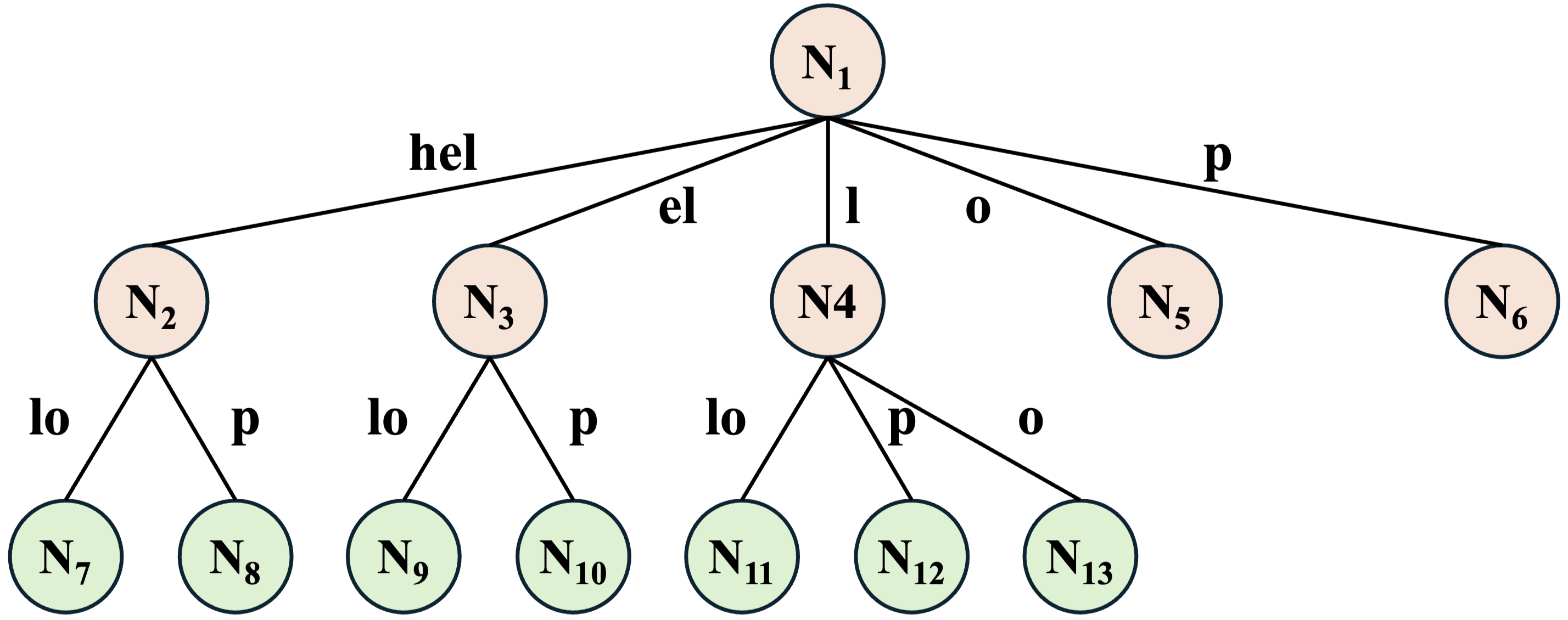}
\caption{Suffix tree construction using "hello" and "help" as examples.} \label{fig2}
\end{figure}

Building upon foundational work by Chase and Shen \cite{chase2015substring}, Zichen Gui et al. developed an enhanced suffix tree variant that supports concurrent multiple substring queries \cite{gui2024query}. %, achieving both improved computational efficiency and greater query flexibility compared to traditional approaches. 
Figure~\ref{fig2} demonstrates the construction of a suffix tree for substring-SSE using "hello" and "help" as examples, where each leaf node represents a substring index of the input strings.
The construction of suffix trees must satisfy the following conditions.
\begin{itemize}
    \item the number of leaf nodes must equal the string length, 
    \item each node $N$ must have at least two children, 
    \item edges originating from the same node cannot share identical starting characters, 
    \item each node stores string indices (indicating which strings contain the substring) along with their corresponding starting positions.
\end{itemize}

% \subsubsection{EMM}

\subsubsection{Character Equality}
Let $\mathsf{charEq}(a,b)$ denote the length of identical character sequences between strings $a$ and $b$. 
In suffix tree, string $s$ can be represented as combinations of integer tokens. For instance, given $s_1=``ell"$ and $s_2=``elp"$, we have $\mathsf{charEq}(q_1,q_2)=2$. 
Consequently, these strings can be represented by tokens $(1,2,3)$ and $(1,2,4)$ respectively. Thus, the recovery of string reduces to recovering character equality -- specifically, recovering the mapping between alphabet characters and tokens.

\textit{Initial path} refers to the string composed of all characters traversed from the root node to the parent node of the target node, along with the first character from the parent node to the target node.
For an $n$-level suffix tree, let $N$ be a leaf node. 
Then $\mathsf{initpath}(N)$ refers to the concatenation of strings along the first $N-1$ levels of the path to $N$, followed by the first character of the string at the $N$-th level. 
In Figure~\ref{fig2}, the initial path of leaf node $N_{11}$ is $``ll"$, while the initial path of leaf node $N_{12}$ is $``lp"$.
Thus, recovering initial paths equates to substring recovery, and initial path recovery rate equals to correct string recovery rate, i.e., percentage of strings for which all tokens are correctly guessed.

\subsection{Threat Model}

As outlined in the system model, interactions between clients and servers inevitably introduce leakage. This implies that the \textit{searchable ciphertext} (SC) uploaded by the data owner to the cloud server, the \textit{string queries} submitted by the data user, and the \textit{search results} returned to the user are all susceptible to exposure. Among these, SC in substring-SSE schemes contains the most extensive substring information, making it the primary target for recovery in this study.

The attacker’s objective is thus to reconstruct the token, string, and initial path mappings via observed leakage. In our leakage-abuse attack model, the adversary is assumed to be \textbf{passive}, monitoring protocol executions and acquiring partial leakage from SC, such as character and string frequencies and distributions, which forms a partially known dataset. Using this information, the attacker attempts to recover the $(a, t)$ and $(s, es)$ mappings, ultimately aiming for full SC reconstruction. Scenarios involving partial queries and search results, though beyond the current scope, present meaningful avenues for future research.

To the best of our knowledge, leakage in substring-SSE schemes can be classified into three types, summarized below. Note that when a string “retrieves” certain nodes, it signifies that the substrings of the string traverse those nodes during the search process.

\subsubsection{Prefix pattern leakage}
When examining prefix node indices retrieved by string $s_l$, the system reveals whether the index was previously retrieved by any strings in $(s_1,...,s_{l-1})$. 
     For example, assume that $s_1=(A,C,E), s_2=(A,B,C), s_3=(B,D,E), s_4=(C,D,E)$, where $A,...,E$ are prefix nodes in suffix trees, it is obvious that $l=4$. The leakage of $s_4$ can be represented as $\mathbf{L_1}=l\times n_i$ matrix below, where $n_i$ is the nodes sum retrieved by $s_l$.
     If $\mathbf{L_1}(i, j) = 1$, it means that $s_i$ has visited $n_j$, otherwise it is 0.
\subsubsection{Leaf node intersection pattern leakage}
For each leaf node index retrieved by string $s_l$, the system reveals whether the node was also retrieved by any prior strings in $(s_1,...,s_{l-1})$.  
    Assume that $s_l$ retrieves 4 leaf nodes, the indices are $[1,2,3,4]$, which is randomized to $[3,1,4,2]$ by random permutation function $r_2:[m_j]\rightarrow [m_j]$. 
    For string $s_1=(3,4)$ and $s_2=(1,2)$, the leakage can be represented as $\mathbf{L_2}=l\times n_i$ matrix below.
%     $$L_1 = \bordermatrix{
%     & C & D & E \cr
%     s_1 & 1 & 0 & 0 \cr
%     s_2 & 1 & 0 & 0 \cr
%     s_3 & 0 & 1 & 1 \cr
%     s_4 & 1 & 1 & 1 
% }, \quad 
% L_2 = \bordermatrix{
%     & r_2(1) & r_2(2) & r_2(3) & r_2(4) \cr
%     s_1 & 1 & 0 & 1 & 0 \cr
%     s_2 & 0 & 1 & 0 & 1 \cr
% }.$$

\[
\mathbf{L_1}=
\begin{bNiceMatrix}[
    first-row,
    first-col,
    % code-for-first-row = \color{blue},
    % code-for-first-col = \color{blue}
]
        & C & D & E \\
s_1    & 1 & 0 & 0 \\
s_2    & 1 & 0 & 0 \\
s_3    & 0 & 1 & 1 \\
s_4    & 1 & 1 & 1 
\end{bNiceMatrix}
\hspace{2em}  % 可调整间距（如1em/3em等）
\mathbf{L_2}=
\begin{bNiceMatrix}[
    first-row,
    first-col,
    % code-for-first-row = \color{blue},
    % code-for-first-col = \color{blue}
]
        & r_2(1) & r_2(2) & r_2(3) & r_2(4) \\
s_1    & 1 & 0 & 1 & 0 \\
s_2    & 0 & 1 & 0 & 1 
\end{bNiceMatrix}
\]

    \subsubsection{Index intersection pattern leakage}
    When examining each index retrieved by string $s_l$, the system reveals whether the index was previously retrieved by any strings $(s_1,...,s_{l-1})$. 
    This leakage is similar in principle to the leaf node intersection pattern leakage, which will not be discussed here.

    To simplify and unify the leakage characterization, we reduce these observations to character uniqueness leakage. 
    Specifically, the distribution patterns of substrings in the suffix tree reveal uniquely identifiable character information. 
    Since all three leakage patterns can be represented in matrix form, they can ultimately be reduced to either unique row/column permutations or summation uniqueness in the matrix representation.

%这三种泄漏模式都可以用矩阵来表示，因此可以归约为字符的唯一性泄漏以及矩阵行列排列或者和的唯一性泄漏。

% \begin{enumerate}
%     \item \textbf{Prefix pattern leakage}.     \item \textbf{Leaf node intersection pattern leakage}. For each leaf node index retrieved by query $q_l$, the system reveals whether the node was also retrieved by any prior queries in $(q_1,...,q_{l-1})$.  
%     Assume that $q_l$ retrieves 4 leaf nodes, the indices are $[1,2,3,4]$, which is randomized to $[3,1,4,2]$ by random permutation function $r_2:[m_j]\rightarrow [m_j]$. 
%     For query $q_1=(3,4)$ and $q_2=(1,2)$, the leakage can be represented as $L_3=l\times n_i$ matrix below.
%     $$L_2 = \bordermatrix{
%     & C & D & E \cr
%     q_1 & 1 & 0 & 0 \cr
%     q_2 & 1 & 0 & 0 \cr
%     q_3 & 0 & 1 & 1 \cr
%     q_4 & 1 & 1 & 1 
% }, \quad 
% L_3 = \bordermatrix{
%     & r_2(1) & r_2(2) & r_2(3) & r_2(4) \cr
%     q_1 & 1 & 0 & 1 & 0 \cr
%     q_2 & 0 & 1 & 0 & 1 \cr
% }.$$
%     \item \textbf{Index intersection pattern leakage}. When examining each string index retrieved by query $q_l$, the system reveals whether the index was previously retrieved by any queries $(q_1,...,q_{l-1})$. 
%     This leakage is similar in principle to the leaf node intersection pattern leakage, which will not be discussed here.
% \end{enumerate}

\section{Our Attack}\label{sec:attack}
\subsection{Notations}
We use $s$,$a$,$es$,$t$ to respectively denote a string, an alphabet, an encrypted string, and a token. 
We use $s_i$ to denote a specific $\textsf{string}_i$, and use $a_i$,$es_i$, and $t_i$ similarly. 
Likewise, let $S=\{s_1, ... ,s_n\}$ denote the set of plaintext strings of a user, $A=\{a_1, ... ,a_m\}$ denote the set of alphabets appear in $S$, which is same for $E=\{es_1, ... ,es_n\}$ and $T=\{t_1, ... ,t_n\}$.

Note that $s(\textsf{reps}.a)$ is indexed independently from $es(\textsf{reps}.t)$. 
In other words, $es_i$ may not derive from $s_i$ and $t_i$ might not correspond to $a_i$ despite shared indices. 
Additionally, we use $(es, s)$ to denote a mapping between an encrypted string and the corresponding plaintext string, and use $(t, a)$ to denote a mapping between a token and the corresponding alphabet.

% \subsection{Knowledge of Attacker}

% The attacker aims to recover the token, string query, and the initial path through the leakage.

% The system model primarily generates access pattern leakage, with the most exploitable types being prefix pattern leakage, leaf node intersection pattern leakage, and index intersection leakage.

\subsection{Attack Description}
Based on the previous discussion, the goal of the adversary is to recover SC, which can be represented as a mapping between the encrypted string and the plaintext string, denoted as $(es,s)$, and a mapping between the alphabet token and the alphabet, denoted as $(t,a)$.

Let $S_k=\{s_1, \ldots, s_n\}$ denote the complete set of strings contained in document $k$, and $A=\{a_1, \ldots, a_m\}$ represent all distinct characters in strings, where $n$ is the total number of strings and $m$ is the total number of characters appeared in strings. 
Under the given adversarial assumptions, the attacker possesses partial knowledge of the database:
\begin{itemize}
    \item A subset of strings $S_k^{'}=\{s_{y_1}, \ldots, s_{y_{n'}}\} \subseteq S_k$;
    \item The corresponding character set $A'_k=\{a_{x_1}, \ldots, a_{x_{m'}}\} \subseteq A$ contained in $S'_k$.
\end{itemize}
where $\{y_1, \ldots, y_{n'}\} \subset [n]$ and $\{x_1, \ldots, x_{m'}\} \subset [m]$ denote the index sets of the known strings and characters, respectively.

Based on the system model,  mapping information of the parameters can be represented in the form of matrices below.
\[\mathbf{A}=
\begin{bNiceMatrix}[
    first-row,
    first-col,
    code-for-first-row = \color{blue},
    code-for-first-col = \color{blue}
]
        & s_1    & s_2    & \cdots & s_n     \\
a_1     & A_{1,1} & A_{1,2} & \cdots & A_{1,n}  \\
a_2     & A_{2,1} & A_{2,2} & \cdots & A_{2,n}  \\
\vdots  & \vdots & \vdots & \ddots & \vdots  \\
a_m     & A_{m,1} & A_{m,2} & \cdots & A_{m,n}
\end{bNiceMatrix}
\hspace{2em}  % 可调整间距（如1em/3em等）
\mathbf{B}=
\begin{bNiceMatrix}[
    first-row,
    first-col,
    code-for-first-row = \color{blue},
    code-for-first-col = \color{blue}
]
        & es_1    & es_2    & \cdots & es_n     \\
t_1     & B_{1,1} & B_{1,2} & \cdots & B_{1,n}  \\
t_2     & B_{2,1} & B_{2,2} & \cdots & B_{2,n}  \\
\vdots  & \vdots & \vdots & \ddots & \vdots  \\
t_m     & B_{m,1} & B_{m,2} & \cdots & B_{m,n}
\end{bNiceMatrix}
\]

Matrices $\mathbf{A}$ and $\mathbf{B}$ are both $m \times n$ matrices, where matrix $\mathbf{A}$ represents the correspondence between $S_k$ and the character set $A$, while matrix $\mathbf{B}$ represents the correspondence between the set of ciphertext strings $ES_k$ and the set of tokens $T$ contained in the strings. 
In other words, matrix $\mathbf{B}$ can be regarded as an encrypted version of matrix $\mathbf{A}$. 
If the string $s_{j}$ contains the character $a_{i}$, then $\mathbf{A_{i,j}}$ equals 1; otherwise, it equals 0. 
The same applies to $\mathbf{B_{i,j}}$.
The attacker gains access to $\mathbf{B}$ and partial knowledge of $\mathbf{A}$ through leakage.
\[
\begin{array}{c}
\mathbf{A^{'}} =
\begin{bNiceMatrix}[
    first-row,
    first-col,
    code-for-first-row = \color{blue},
    code-for-first-col = \color{blue}
]
        & s_{y_1}    & s_{y_2}    & \cdots & s_{y_{n^{'}}}     \\
a_{x_1}     & A^{'}_{1,1} & A^{'}_{1,2} & \cdots & A^{'}_{1,n}  \\
a_{x_2}     & A^{'}_{2,1} & A^{'}_{2,2} & \cdots & A^{'}_{2,n}  \\
\vdots  & \vdots & \vdots & \ddots & \vdots  \\
a_{x_{m^{'}}}     & A^{'}_{n,1} & A^{'}_{n,2} & \cdots & A^{'}_{n,n}
\end{bNiceMatrix}
\\[3ex]  % 调整上下间距
\Downarrow  \\[1ex]  % 箭头与下方矩阵的间距
\mathbf{A^{''}} =
\begin{bNiceMatrix}[
    first-row,
    first-col,
    code-for-first-row = \color{blue},
    code-for-first-col = \color{blue}
]
        & s_{y_1}    & s_{y_2}    & \cdots & s_{y_{n^{'}}}     \\
a_{x_1}     & A_{1,1}^{''} & A_{1,2}^{''} & \cdots & A_{1,n}^{''}  \\
a_{x_2}     & A_{2,1}^{''} & A_{2,2}^{''} & \cdots & A_{2,n}^{''}  \\
\vdots  & \vdots & \vdots & \ddots & \vdots  \\
a_{x_{m^{'}}}     & A_{m^{'},1}^{''} & A_{m^{'},2}^{''} & \cdots & A_{m^{'},n^{'}}^{''}  \\
a_{x_{m^{'}+1}}     & A_{m^{'}+1,1}^{''}=0 & A_{m^{'}+1,2}^{''}=0 & \cdots & A_{m^{'}+1,n^{'}}^{''}=0  \\
\vdots  & \vdots & \vdots & \cdots & \vdots  \\
a_{x_{m^{''}}}     & A_{m^{''},1}^{''}=0 & A_{m^{''},2}^{''}=0 & \cdots & A_{m^{''},n^{'}}^{''}=0  \\
\end{bNiceMatrix}
\end{array}
\]
Since the attacker can obtain $S_k^{'}=\{s_{y_1}, \ldots, s_{y^{'}_{n}}\}$ and its corresponding character set $A'_k=\{a_{x_1}, \ldots, a_{x_{m'}}\}$, the attacker's knowledge can be represented as an $m \times n$ matrix $\mathbf{A'}$. 
Similarly, if the string $s_{y_j}$ contains the character $a_{x_i}$, then $\mathbf{A'_{i,j}}$ equals 1; otherwise, it equals 0.

Based on the knowledge of matrix $\mathbf{A'}$, the attacker first extends the number of rows of $\mathbf{A'}$ to $m$ by setting $\mathbf{A''}_{i,j} = 0$ for $i \in \{m'+1, \ldots, m''\}$ and $j \in \{1, \ldots, n'\}$, and obtains the $m \times n'$ matrix $\mathbf{A''}$. 
    Specifically, $\{x_1, \ldots, x_m\} = \{x_1, \ldots, x_{m'}\} \cup \{x_{m'+1}, \ldots, x_{m''}\}$, where $x_i$ uniquely corresponds to an alphabet character while the attacker only knows $\{x_1, \ldots, x_{m'}\}$, and $\{x_{m'+1}, \ldots, x_{m''}\}$ are unknown to the attacker.
    Now the number of rows of matrix $\mathbf{B}$ and $\mathbf{A''}$ are the same.

The attacker infers matrices $\mathbf{M}$ and $\mathbf{M^{'}}$, where \(\mathbf{M_{i,j}}  \) denotes the number of shared characters between encrypted strings \( es_i \) and \( es_j \), and \(\mathbf{M'_{i,j}}  \) denotes the number of shared characters between known plaintext strings \( s_{y_i} \) and \( s_{y_j} \).
\[\mathbf{M}=
\begin{bNiceMatrix}[
    first-row,
    first-col,
    code-for-first-row = \color{blue},
    code-for-first-col = \color{blue}
]
        & es_1    & es_2    & \cdots & es_n     \\
es_1     & M_{1,1} & M_{1,2} & \cdots & M_{1,n}  \\
es_2     & M_{2,1} & M_{2,2} & \cdots & M_{2,n}  \\
\vdots  & \vdots & \vdots & \ddots & \vdots  \\
es_n     & M_{n,1} & M_{n,2} & \cdots & M_{n,n}
\end{bNiceMatrix}
\hspace{2em}  % 可调整间距（如1em/3em等）
\mathbf{M^{'}}=
\begin{bNiceMatrix}[
    first-row,
    first-col,
    code-for-first-row = \color{blue},
    code-for-first-col = \color{blue}
]
        & s_{y_1}    & s_{y_2}    & \cdots & s_{y_{n^{'}}}     \\
s_{y_1}     & M_{1,1}^{'} & M_{1,2}^{'} & \cdots & M_{1,n}^{'}  \\
s_{y_2}     & M_{2,1}^{'} & M_{2,2}^{'} & \cdots & M_{2,n}^{'}  \\
\vdots  & \vdots & \vdots & \ddots & \vdots  \\
s_{y_{n^{'}}}     & M_{n,1}^{'} & M_{n,2}^{'} & \cdots & M_{n,n}^{'}
\end{bNiceMatrix}
\]

It is not difficult to see that an attacker can easily obtain the matrices $\mathbf{B}$, $\mathbf{A'}$, $\mathbf{A''}$ ,$\mathbf{M}$, and $\mathbf{M'}$.
Observe that each encrypted string uniquely corresponds to a plaintext string, there exists a subset $Set_{col} \subset \{es_1, \ldots, es_n\}$ such that $\{f_1(s_{y_1}), \ldots, f_1(s_{y_{n'}})\} = Set_{col}$, where $f_1$ is a function, therefore for each column of $\mathbf{A''}$, there exists a matching column in $\mathbf{B}$. 
%Similarly, note that each token uniquely corresponds to an alphabet character, there exists a subset $Set_{row} \subset \{t_1, \ldots, t_m\}$ such that $\{f_2(a_{x_1}), \ldots, f_2(a_{x_{m'}})\} = Set_{row}$, where $f_2$ is a function, hence for each row of $\mathbf{A'}$, there exists a matching row in $\mathbf{B}$.

Then the attacker proceeds to carry out the following steps of the attack.

\subsubsection{Unique column-sum mapping}
This step aims to find \textcolor{blue}{$(es,s)$} mapping through unique column-sum mapping. 
    Based on the above matrices, the attacker can establish a unique column sum mapping between these two matrices. 
    For matrix $\mathbf{B}$, if the column sum $Sum_{k_B}$ of column $es_k$ is unique within the set of column sums $\{Sum_{j_B}\}_{j \in [n]}$, then there must exist a corresponding column $s_{y_{k'}}$ in matrix $\mathbf{A '}$ such that $Sum_{k_B} = Sum_{k'_{A '}}$. 
    Consequently, the attacker can recover the mapping pair $(es_k, s_{y_{k'}})$.
    The complete procedure is presented in Algorithm \ref{alg:column_mapping_io} in the Appendix.

\subsubsection{Occurrence matrix mapping}
This step aims to recover more \textcolor{blue}{$(es,s)$} mapping through occurrence matrices mapping.
    Leveraging the known column mappings obtained from previous step and matrices $\mathbf{M}$, $\mathbf{M'}$, the attacker can recover previously unrecovered column mappings. 
    The mapping relationship between $\mathbf{M}$ and $\mathbf{M'}$ indicates that when $\mathbf{M_{i,j}}$ equals $\mathbf{M'_{i',j'}}$, this implies $es_i$ is the encrypted version of $s_i$ and $es_j$ is the encrypted version of $s_j$. 
    For a known mapping pair $(es_k,s_{y_{k'}})$ and an unmapped $s_{y_{j'}}$, if there exists exactly one $es_j$ satisfying both $\mathbf{M_{j,k}} = \mathbf{M'_{j',k'}}$ and the sum of column $j'$ in $\mathbf{A''}$ equaling the sum of column $j$ in $\mathbf{B}$, then a new mapping $(es_j,s_{y_{j'}})$ can be derived. 
    The complete procedure is presented in Algorithm \ref{alg:occurrence_mapping} in Appendix.

\subsubsection{Unique row mapping}
This step aims to recover \textcolor{blue}{$(t,a)$} mapping through unique row mapping.
    Let $\mathbf{S_c} = \{(es_{j_1}, s_{y_{j'_1}}), \ldots, (es_{j_t}, s_{y_{j'_t}})\}$ denote the recovered $(es, s)$ mappings from previous steps, where $\{j_1, \ldots, j_t\} \subset [n]$ and $\{y_{j'_1}, \ldots, y_{j'_t}\} \subseteq [n']$. Based on this construction, the attacker partitions matrices $\mathbf{B}$ and $\mathbf{A''}$, then rearrange their columns according to the order of $(es_{j_1}, es_{j_2}, \ldots, es_{j_t})$ and $(s_{y_{j'_1}}, s_{y_{j'_2}}, \ldots, s_{y_{j'_t}})$ respectively, obtaining submatrices $\mathbf{B_c}$ and $\mathbf{A''_c}$ with column-wise correspondence. 
\[
\mathbf{B_c}=
\begin{bNiceMatrix}[
    first-row,
    first-col,
    code-for-first-row = \color{blue},
    code-for-first-col = \color{blue}
]
        & es_{j_1}    & es_{j_2}    & \cdots & es_{j_t}     \\
t_1     & B_{1,j_1} & B_{1,j_2} & \cdots & B_{1,j_t}  \\
t_2     & B_{2,j_1} & B_{2,j_2} & \cdots & B_{2,j_t}  \\
\vdots  & \vdots & \vdots & \ddots & \vdots  \\
\color{red}t_m & \color{red}B_{m,j_1} & \color{red}B_{m,j_2} & \color{red}\cdots & \color{red}B_{m,j_t}
\end{bNiceMatrix}
\hspace{2em}
\mathbf{A^{''}_c}=
\begin{bNiceMatrix}[
    first-row,
    first-col,
    code-for-first-row = \color{blue},
    code-for-first-col = \color{blue}
]
        & s_{y_{j^{'}_1}}    & s_{y_{j^{'}_2}}    & \cdots & s_{y_{j^{'}_t}}     \\
a_1     & A_{1,j^{'}_1}^{''} & A_{1,j^{'}_2}^{''} & \cdots & A_{1,j^{'}_t}^{''}  \\
\color{red}a_2 & \color{red}A_{2,j^{'}_1}^{''} & \color{red}A_{2,j^{'}_2}^{''} & \color{red}\cdots & \color{red}A_{2,j^{'}_t}^{''} \\
\vdots  & \vdots & \vdots & \ddots & \vdots  \\
a_m     & A_{m,j^{'}_1}^{''} & A_{m,j^{'}_2}^{''} & \cdots & A_{m,j^{'}_t}^{''}
\end{bNiceMatrix}
\]
    Furthermore, if the $i$-th row bit-string pattern in $\mathbf{B_c}$ is unique, this implies its uniqueness in both $\mathbf{B}$ and the corresponding rows of $\mathbf{A''_c}$ and $\mathbf{A''}$. Consequently, when a unique bit-string pattern in row $i$ of $\mathbf{B_c}$ matches row $i'$ of $\mathbf{A''_c}$, we can deduce that $t_i$ corresponds to $a_{i'}$.
    Taking the two matrices above as examples, if the $m$-th row of matrix $\mathbf{B_c}$ matches the $2$-nd row of matrix $\mathbf{A''_c}$, we can derive the $(t_m, a_2)$ mapping.
    The complete procedure is presented in Algorithm \ref{alg:row_mapping_io} in the Appendix.

\subsubsection{Unique column mapping}
    This step aims to recover more \textcolor{blue}{$(es,s)$} mapping through unique column mapping. 
    Let $S_r=\{(t_{i_1},a_{x_{i'_1}}), \ldots, (t_{i_t},a_{x_{i'_t}})\}$ be recovered $(t,a)$ mappings from the third step, where $\{i_1, \ldots, i_t, i'_1, \ldots, i'_t\} \subseteq [m]$. The attacker rearranges the rows of matrix $\mathbf{B}$ into a submatrix $\mathbf{B}_r$ according to $(t_{i_1}, \ldots, t_{i_t})$, and the rows of matrix $\mathbf{A}''$ into a submatrix $\mathbf{A}''_r$ according to $(a_{x_{i'_1}}, \ldots, a_{x_{i'_t}})$.
\[
\mathbf{B_r}=
\begin{bNiceMatrix}[
    first-row,
    first-col,
    code-for-first-row = \color{blue},
    code-for-first-col = \color{blue}
]
        & es_{1}    & \color{red}es_{2}    & \cdots & es_{n}     \\
t_{i_1}     & B_{i_1,1} & \color{red}B_{i_1,2} & \cdots & B_{i_1,n}  \\
t_{i_2}     & B_{i_2,1} & \color{red}B_{i_2,2} & \cdots & B_{i_2,n}  \\
\vdots  & \vdots & \color{red}\vdots & \ddots & \vdots  \\
t_{i_t}     & B_{i_t,1} & \color{red}B_{i_t,2} & \cdots & B_{i_t,n}
\end{bNiceMatrix}
\hspace{2em}
\mathbf{A^{''}_r}=
\begin{bNiceMatrix}[
    first-row,
    first-col,
    code-for-first-row = \color{blue},
    code-for-first-col = \color{blue}
]
        & s_{y_{1}}    & s_{y_2}    & \cdots & s_{y_{n^{'}}}     \\
a_{x_{i_1^{'}}}     & \color{red}A_{i_1^{'},1}^{''} & A_{i_1^{'},2}^{''} & \cdots & A_{i_1^{'},n^{'}}^{''}  \\
a_{x_{i_2^{'}}}     & \color{red}A_{i_2^{'},1}^{''} & A_{i_2^{'},2}^{''} & \cdots & A_{i_2^{'},n^{'}}^{''}   \\
\vdots  & \color{red}\vdots & \vdots & \ddots & \vdots  \\
a_{x_{i_t^{'}}}     & \color{red}A_{i_t^{'},1}^{''} & A_{i_t^{'},2}^{''} & \cdots & A_{i_t^{'},n^{'}}^{''} 
\end{bNiceMatrix}
\]
    Thus, the rows of $\mathbf{B_r}$ and $\mathbf{A''_r}$ are in one-to-one correspondence. Furthermore, if the $j$-th column bit-string of $\mathbf{B_r}$ is unique, then this column is also unique in $\mathbf{B}$, and similarly for $\mathbf{A''_r}$ and $\mathbf{A''}$. 
    Therefore, if the unique bit-string in the $j$-th column of $\mathbf{B_r}$ matches the $j'$-th column of $\mathbf{A''_r}$, we can conclude that the string corresponding to $es_j$ is $s_{y_{j'}}$.
    Taking the two matrices above as examples, if the $2$-nd column of matrix $\mathbf{B_r}$ matches the $1$-st row of matrix $\mathbf{A''_r}$, we can derive the $(es_2, s_{y_1})$ mapping.
    The complete procedure is presented in Algorithm \ref{alg:unique_column_mapping} in the Appendix.

\subsubsection{Iterative recovery}
    This step aims to recover more \textcolor{blue}{$(es,s)$} mapping through column-sum mapping of recomputed matrices. 
    Let $\mathbf{V_{B_j}}$ denote the column-sum vector of matrix $\mathbf{B}$, such that $\mathbf{V_{B_j}} = [\mathrm{Sum}_{1B}, \mathrm{Sum}_{2B}, \ldots, \mathrm{Sum}_{nB}]$ and $\mathbf{V_{A''_{j'}}} = [\mathrm{Sum}_{1A''}, \mathrm{Sum}_{2A''}, \ldots, \mathrm{Sum}_{nA''}]$ represent the column-sum vector of matrix $\mathbf{A''}$. 
    First, all matched elements in matrices $\mathbf{B}$ and $\mathbf{A''}$ are set to 0. Then, for each unmatched column $j$ in $\mathbf{B}$, we recompute the value $\mathrm{Sum}_{jB}$ and update it in $\mathbf{V_{B_j}}$. 
    The same procedure applies to $\mathbf{V_{A''_{j'}}}$.
    Then, based on the computed $\mathbf{V_{B_j}}$ and $\mathbf{V_{A''_{j'}}}$, the attacker can derive mapping $(es_j,s_{y_{j^{'}}})$ if $Sum_{jB}=Sum_{j^{'}A^{''}}$, where $j$ and $y_{j^{'}}$ are the indices of the umapped column in $\mathbf{B}$ and $\mathbf{A''}$ respectively.
    The complete procedure is presented in Algorithm \ref{alg:iterative_recovery} in the Appendix.

\section{Experimental Evaluation}\label{sec:evaluation}
% Below is the experimental evaluation section of this paper. 
% Our attack represents the first known dataset-based attack on substring-SSE schemes under a partial knowledge setting. 
% Specifically, this attack targets the suffix tree data structure proposed in the Chase-Shen scheme, focusing on the recovery of $(a, t)$ mappings and $(s, es)$ pairs.

\subsection{Setup}
This attack is based on string leakage, meaning the attack still works even if the known data is fragmented (e.g., random text snippets). However, for ease of quantification, this experiment uses partially complete documents to simulate the attacker’s prior knowledge.
We use the Enron email dataset, including 30109 emails with over 1,000,000 strings, which has been widely adopted in SSE literature \cite{bag2023two,gui2024swissse,naveed2015fallacy,nie2024query,ning2021leap,oya2021hiding}. 
The Enron dataset provides a cryptographically meaningful testbed for substring-SSE evaluation due to its natural language characteristics and structural properties. 
Its heterogeneous string lengths, non-uniform character distribution (covering 94 symbols, e.g., a, A, ..., 1, <, !.), and semantically correlated substrings accurately model real-world text patterns that affect search leakage profiles. 
These features make it particularly suitable for analyzing both entropy characteristics and semantic dependencies in practical substring search scenarios.

Following the methodology in literature \cite{cash2015leakage,zhang2016all}, we generate a set of stop words (e.g., "the," "to," "of," etc.) to extract strings. 
Additionally, to validate the effectiveness of character recovery in our scheme, we first shuffle all 94 distinct characters (including uppercase and lowercase letters, digits, and punctuation marks) appearing in the emails and randomly map them to three-digit integers\cite{noor2022simple}. 
These 94 randomly mapped three-digit integers can then be treated as ciphertext representations of the characters.

\subsection{Experimental Design}
% In effectiveness test, the experimental procedure is as follows. First, we randomly select 5,000 strings from the emails as experimental samples and partition them into the attacker's partially known dataset at rates of "1\%, 5\%, 10\%, ..., 100\%." 
% The dataset includes plaintext strings and their corresponding characters. 
% The attacker then preprocesses these datasets and represents parameter relationships in matrix form, subsequently recovering the $(a, t)$ and $(s, es)$ mappings through matrix transformations. 
In the effectiveness evaluation, the experimental procedure is structured as follows. First, we randomly sample 5,000 strings from the email corpus as experimental data and partition them into the attacker's partially known dataset at ratios of 1%, 5%, 10%, \ldots, 100%.
The dataset comprises plaintext strings along with their corresponding ciphertext characters.
The attacker then preprocesses these data and represents parametric relationships in matrix form, subsequently reconstructing the $(a, t)$ and $(s, es)$ mappings via matrix transformations.

Robustness testing is conducted under the assumption that the attacker possesses knowledge of 10\% (500 auxiliary samples) of a 5,000-sample dataset. 
The attacker attempts to recover the $(a, t)$ mappings and then recover $(s,es)$ mappings, thereby reconstructing the remaining strings in each dataset based on 500 strings.
By progressively increasing the scale of the target database from 1000 to 30000, this test evaluates whether the attack recovery rates of various parameters decrease as the target database size expands.

\subsection{Results}
The experimental results are presented from two perspectives: \textit{effectiveness} and \textit{robustness}. 
\textit{Effectiveness} refers to the recovery rate of the attack parameters, while \textit{robustness} indicates that the recovery performance remains stable and does not degrade as the number of ciphertext strings increases.

\subsubsection{Effectiveness}
\begin{figure}
\centering
\includegraphics[width=0.9\textwidth]{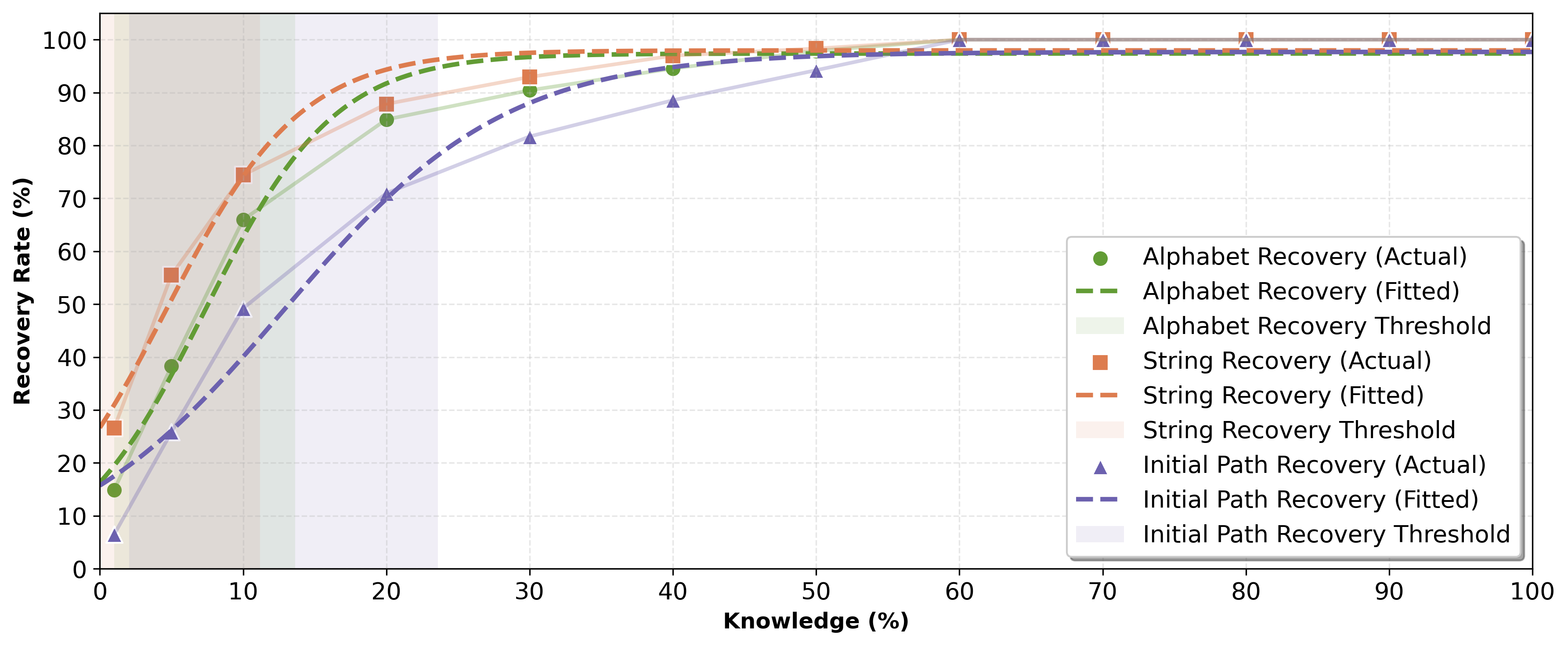}
\caption{Recovery rate of character, string and initial path under varied knowledge set.} \label{fig3}
\end{figure}

Effectiveness test involves simulating knowledge-based attacks against 5,000 strings, followed by nonlinear least-squares logistic regression \cite{nick2007logistic} to determine parameter evolution characteristics and detection thresholds, where $L$ denotes the \textit{asymptotic upper bound} and $k$ governs the \textit{growth rate}. 
Table~\ref{tab:combined_recovery} presents the combined recovery of alphabet, string and initial path with respect to knowledge set in detail.
Figure~\ref{fig3} shows the recovery rates of the attack on the alphabetic characters, strings, and initial paths. 
The x-axis represents the dataset knowledge, ranging from 1\% to 100\%.
The y-axis represents the recovery rate, which is the rate of recovered characters, strings, and initial paths to the total number of characters, strings, and initial paths in the dataset.

\begin{table}[h]
\centering
\caption{Combined recovery statistics with respect to different knowledge sets.}
\label{tab:combined_recovery}
\begin{tabular}{lcccccc}
\toprule
 & \multicolumn{2}{c}{\textbf{Recovered Alphabet}} & \multicolumn{2}{c}{\textbf{Recovered String}} & \multicolumn{2}{c}{\textbf{Initial Path}} \\
\cmidrule(lr){2-3} \cmidrule(lr){4-5} \cmidrule(lr){6-7}
\textbf{Knowledge (\%)} & \textbf{Count} & \textbf{Rate (\%)} & \textbf{Count} & \textbf{Rate (\%)} & \textbf{Count} & \textbf{Rate (\%)} \\
\midrule
1.0  & 14  & 14.89 & 1329 & 26.58 & 324 & 6.48 \\
5.0   & 36  & 38.30 & 2775 & 55.5 & 1292 & 25.84 \\
10.0  & 62  & 65.96 & 3721 & 74.42 & 2461 & 49.22 \\
20.0  & 80 & 84.91 & 4392 & 87.84 & 3549 & 70.98 \\
30.0  & 85 & 90.43 & 4674 & 92.94 & 4085 & 81.70 \\
40.0  & 89  & 94.55 & 4849 & 96.98 & 4427 & 88.54 \\
50.0  & 92  & 97.87 & 4916 & 98.32 & 4711 & 94.22 \\
60.0  & 94  & 100.00 & 5000 & 100.00 & 5000 & 100.00 \\
% 70.0  & 94  & 100.00 & 5000 & 100.00 & 5000 & 100.00 \\
% 80.0  & 94  & 100.00 & 5000 & 100.00 & 5000 & 100.00 \\
% 90.0  & 94  & 100.00 & 5000 & 100.00 & 5000 & 100.00 \\
100.0 & 94  & 100.00 & 5000 & 100.00 & 5000 & 100.00 \\
\bottomrule
\end{tabular}
\end{table}

%这一段是要改的
% This experiment conducts a systematic evaluation on 5,000 sample strings to analyze the performance of character recovery, string recovery, and exact string recovery under varying proportions of known data.
% The results indicate that character recovery exhibits a clear threshold effect---reaching 65.96\% at 10\% knowledge and achieving full recovery once the known proportion exceeds 60\%. 
% String recovery demonstrates remarkable early-stage effectiveness, achieving 26.58\% success with just 1\% known data and showing exponential growth between 10\%-20\% knowledge intervals (from 74.42\% to 87.84\%).
% Exact string recovery shows consistent linear improvement, progressing from 15.48\% to 53.22\% between 1\%-10\% knowledge, and ultimately reaching full recovery at 60\% knowledge.
% These findings confirm the enhanced early-stage effectiveness of the proposed method while maintaining terminal convergence, demonstrating that even minimal (1\%) known data enables significant (26.58\%) path recovery in substring-SSE schemes.

The experimental results of the attack demonstrate a characteristic S-shaped recovery pattern across alphabet, string, and initial path recovery.
The fitted parameters reveal that alphabet recovery achieves the steepest growth ($k = 0.2195 \pm 0.0311$) with a 50\% recovery threshold at 7.31\% knowledge, while string recovery shows comparable asymptotic accuracy ($L = 97.96\% \pm 1.30\%$) but reaches its midpoint earlier at 4.64\% knowledge. 
Initial path recovery exhibits the most gradual growth ($k = 0.1287 \pm 0.0203$), requiring 12.83\% knowledge to achieve 50\% recovery. 
All three parameters demonstrate high model fidelity ($R^2 > 0.97$), with threshold windows (1.0\%–13.6\% for alphabet, 0.0\%–11.2\% for string, and 2.1\%–23.6\% for initial path) indicating the knowledge ranges where the attack transitions from less effective to highly effective. 
%The consistent asymptotic limits near 97\%–98\% before final convergence at 60\% knowledge suggest a small systematic underestimation in the logistic model, possibly due to late-stage synergistic effects between recovery mechanisms.

\subsubsection{Robustness}
\begin{figure}
\centering
\includegraphics[width=0.9\textwidth]{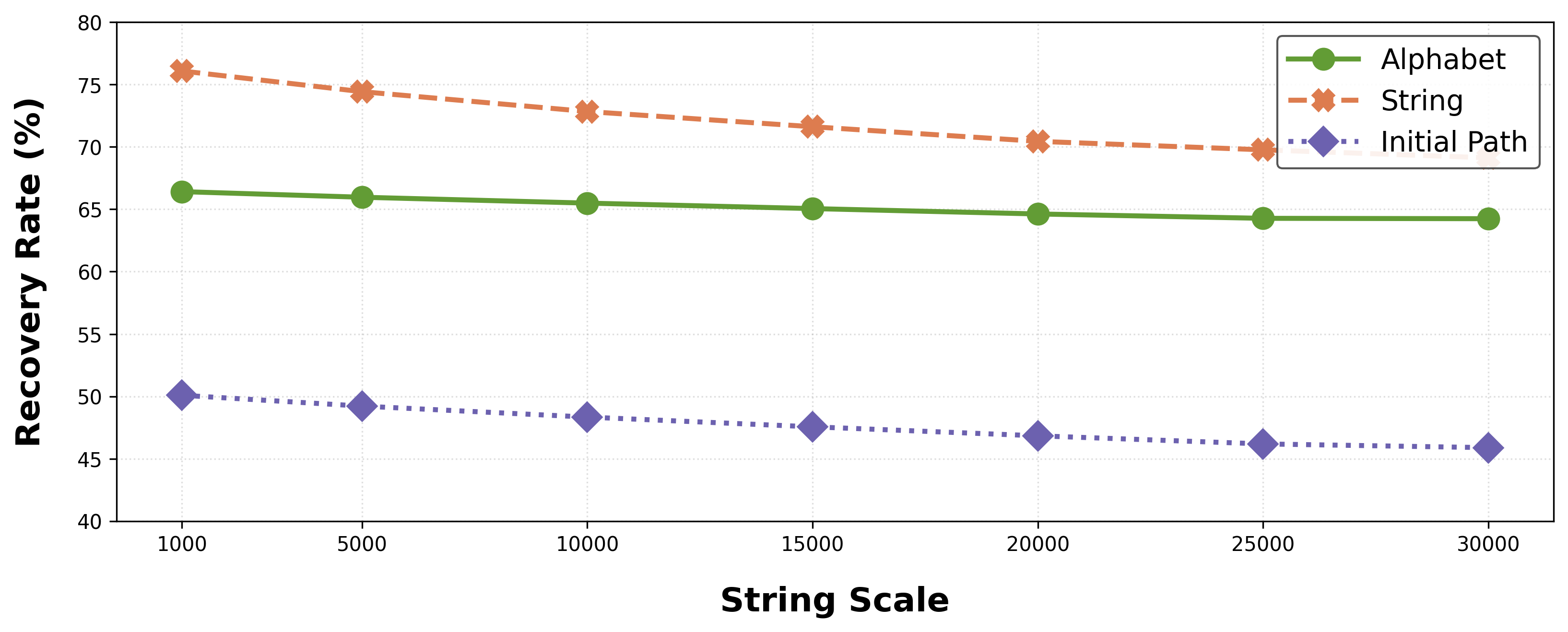}
\caption{Recovery accuracy of character, string and initial path under varied string scale.} \label{fig4}
\end{figure}

%Figure~\ref{fig4} illustrates the robustness of the attack, showing the recovery rates of alphabetic characters, strings, and initial paths as the target database size increases from 1,000 to 30,000.
The robustness test is conducted by using datasets containing 1,000, 5,000, ..., 30,000 strings respectively. Since the effectiveness test demonstrated that the recovery rate exhibits the highest instability when the attacker knows approximately 10\% (i.e., 500 strings) of a 5,000-string dataset, the robustness experiment assumes that each attacker possesses prior knowledge of 500 strings. 
Based on these 500 strings, the attacker attempts to recover the $(a, t)$ mapping and subsequently recover the remaining strings in each dataset, thereby examining the relationship between the recovery rate and the scale of the dataset.
Results are demonstrated in Figure~\ref{fig4} and Table~\ref{tab:Training_sizes}. 
The x-axis represents the string scale.
The y-axis represents the recovery rate.

\begin{table}[htbp]
    \centering
    \caption{Recovery rate with respect to different string scales.}\label{tab:Training_sizes}
    \fontsize{10}{12}\selectfont
    \begin{tabular}{>{\raggedright}m{3.2cm}*{7}{>{\centering\arraybackslash}m{1.1cm}}}
        \toprule
        \diagbox[width=6.5em, innerleftsep=2pt, innerrightsep=-2pt]{Rate(\%)}{Scale} & 
        1000 & 5000 & 10000 & 15000 & 20000 & 25000 & 30000 \\
        \midrule
        Alphabet & 66.41 & 65.96 & 65.49 & 65.05 & 64.62 & 64.27 & 64.24 \\
        String & 76.08 & 74.42 & 72.83 & 71.61 & 70.44 & 69.76 & 69.12 \\
        Initial Path & 50.09 & 49.22 & 48.34 & 47.56 & 46.84 & 46.19 & 45.89 \\
       \bottomrule
    \end{tabular}
    \label{tab:Training_sizes_smoothed}
\end{table}

%这一段是要改的
Experimental results demonstrate strong robustness, with only marginal degradation in recovery rates as the dataset scales from 1,000 to 30,000 strings. 
The alphabet recovery rate remains highly stable, declining slightly from 66.41\% to 64.24\%, while the string recovery rate gradually decreases from 76.08\% to 69.12\%. 
Similarly, initial path recovery moderates from 50.09\% to 45.89\%. 
Notably, the rate of decay slows significantly at larger scales—particularly beyond 20,000 strings—indicating resilience against dataset expansion. 
This scale-invariant behavior suggests that the underlying pattern recognition and mapping techniques are largely unaffected by data volume, supporting the method’s practical viability in real-world deployments with varying dataset sizes.

\subsection{Comparison with Existing Work}
To the best of our knowledge, the only existing attack on substring-SSE schemes is the one proposed by Zichen Gui et al. \cite{gui2024query}.
A comparation shows in Table~\ref{tab:comparison}.

\begin{table}[htbp]
\centering
\renewcommand{\arraystretch}{1.3}
\caption{Comparison between Zichen Gui et al.\cite{gui2024query} and Our Work}\label{tab:comparison}
\begin{tabular}{@{}l l c c@{}}
\toprule
\textbf{Schemes} & & \textbf{Zichen Gui et al. \cite{gui2024query}} & \textbf{Our Work} \\
\midrule
\textbf{Data Generation Method} & & IID Assumption & Partially Known \\
\textbf{Tuned Parameters} & & $\epsilon = 7$, $t = 3$ & N/A \\
\textbf{Scale Robustness} & & $<$5\% fluctuation & $<$5\% fluctuation \\
\midrule
\multirow{3}{*}{\makecell[l]{\textbf{Max Recovery Rate}}} 
& Alphabet & 60.10\% & 97.87\% \\
& String & 63.60\% & 98.32\% \\
& Initial Path & 66.30\% & 94.22\% \\
\bottomrule
\end{tabular}
\end{table}

The experiments conducted by Zichen Gui et al. utilized an independent and identically distributed (IID) dataset of equal scale to the target as prior knowledge, specifically, \textbf{50\%} of the data served as \textbf{auxiliary knowledge} while the remaining \textbf{50\%} constituted the \textbf{target dataset}. 
By tuning parameters candidate set expansion $\epsilon$ and trimming threshold $t$, their attack achieved up to 60.1\% character recovery and 63.6\% query recovery rates on Wikipedia and genome datasets, with a notable finding that short queries enhanced long-query recovery by 13.1\%. 
Notably, the attack efficacy remained invariant to dataset scale. 

In contrast, our study innovatively adopted the \textbf{partially known dataset} assumption on Enron email data, pioneering matrix transformation for parameter recovery in substring-SSE scenarios. 
Logistic regression modeling ($R^2>0.97$) revealed an S-shaped recovery pattern, where merely 7.31\% prior knowledge sufficed for 50\% alphabet recovery, reaching full recovery at 60\% knowledge. 
Moreover, our method demonstrated superior threshold effects and scale robustness, with less than 5\% recovery rate degradation across varying dataset sizes.

% The table summarizes the differences between the two methods in terms of experimental approaches, as well as the highest recovery rate achieved by the scheme proposed by Zichen Gui et al. with 50\% IID dataset and the recovery rates of our method when 50\% of the dataset is known.

\section{Conclusion and Future Work}\label{sec:conclusion}
This paper presents a comprehensive analysis and practical attack on substring-SSE schemes, demonstrating significant vulnerabilities even under partial knowledge conditions. We first formalize a generic substring-SSE architecture and analyze leakage patterns in existing schemes. 
Building upon Zichen Gui et al.'s framework \cite{gui2024query}, we propose an enhanced leakage-abuse attack leveraging matrix-based correlation techniques to recover encrypted data efficiently. 
Experimental validation on the Enron dataset confirms the attack's effectiveness, achieving 100.00\% recovery with 60\% auxiliary knowledge, while maintaining robust performance (degradation <5\%) as dataset size scales to 30,000 strings. 
Notably, the attack succeeds even with minimal prior knowledge (10\%), attaining 65.96\% alphabet and 74.42\% string recovery, thus establishing its practical viability against real-world deployments of substring-SSE systems.

%The current method primarily targets static datasets, and its applicability to dynamic or frequently updated SSE systems requires additional optimizations for incremental computation. 
However, our approach still exhibits several limitations. For instance, it sacrifices computational efficiency to achieve a higher attack success rate, resulting in suboptimal performance. 
Future work will explore how string length affects recovery rates with performance evaluation, particularly the differences between long and short strings. We also aim to study attacks under more limited adversarial knowledge with formal security proofs, such as recovering strings using only partial query information (e.g., 2 out of 200 queries). 
Another important direction involves enhancing and defending against leakage-abuse attacks in substring-SSE, which remains an open challenge for developing effective countermeasures.

\section*{Acknowledgement}
The work is supported by the National Natural Science Foundation of China (No. 62272348).


\begin{thebibliography}{8}
\bibitem{amorim2023leveraging}
Amorim, I. \& Costa, I. Leveraging searchable encryption through homomorphic encryption: A comprehensive analysis. {\em Mathematics}. \textbf{11}, pp. 2948 (2023).

\bibitem{andola2022searchable}
Andola, N., Gahlot, R., Yadav, V., Venkatesan, S. \& Verma, S. Searchable encryption on the cloud: A survey. {\em The Journal of Supercomputing}. pp. 9952-9984 (2022).

\bibitem{bag2023two}
Bag, A., Talapatra, D., Rastogi, A., Patranabis, S. \& Mukhopadhyay, D. Two-in-one-sse: Fast, scalable and storage-efficient searchable symmetric encryption for conjunctive and disjunctive boolean queries. {\em Proceedings on Privacy Enhancing Technologies}. (2023).

\bibitem{bello2021cloud}
Bello, S., Oyedele, L., Akinade, O., Bilal, M., Delgado, J., Akanbi, L., Ajayi, A. \& Owolabi, H. Cloud computing in construction industry: Use cases, benefits and challenges. {\em Automation in Construction}. \textbf{122}, pp. 103441 (2021).

\bibitem{boldyreva2014efficient}
Boldyreva, A. \& Chenette, N. Efficient fuzzy search on encrypted data. {\em International Workshop on Fast Software Encryption}. pp. 613-633 (2014).

\bibitem{cash2015leakage}
Cash, D., Grubbs, P., Perry, J. \& Ristenpart, T. Leakage-abuse attacks against searchable encryption. {\em Proceedings of the 22nd ACM SIGSAC Conference on Computer and Communications Security}. pp. 668-679 (2015).

\bibitem{chase2015substring}
Chase, M. \& Shen, E. Substring-searchable symmetric encryption. {\em Proceedings on Privacy Enhancing Technologies}. (2015).

\bibitem{curtmola2006searchable}
Curtmola, R., Garay, J., Kamara, S. \& Ostrovsky, R. Searchable symmetric encryption: Improved definitions and efficient constructions. {\em Proceedings of the 13th ACM Conference on Computer and Communications Security}. pp. 79-88 (2006).

\bibitem{faber2015rich}
Faber, S., Jarecki, S., Krawczyk, H., Nguyen, Q., Rosu, M. \& Steiner, M. Rich queries on encrypted data: Beyond exact matches. {\em Computer Security - ESORICS 2015: 20th European Symposium on Research in Computer Security, Vienna, Austria, September 21-25, 2015, Proceedings, Part II 20}. pp. 123-145 (2015).

\bibitem{gao2005psist}
Gao, F. \& Zaki, M. PSIST: Indexing protein structures using suffix trees. {\em 2005 IEEE Computational Systems Bioinformatics Conference}. pp. 212-222 (2005).

\bibitem{gui2023rethinking}
Gui, Z., Paterson, K. \& Patranabis, S. Rethinking searchable symmetric encryption. {\em 2023 IEEE Symposium on Security and Privacy (SP)}. pp. 1401-1418 (2023).

\bibitem{gui2024query}
Gui, Z., Paterson, K. \& Patranabis, S. A query reconstruction attack on the Chase-Shen substring-searchable symmetric encryption scheme. {\em Cryptology ePrint Archive}. (2024).

\bibitem{gui2024swissse}
Gui, Z., Paterson, K., Patranabis, S. \& Warinschi, B. SWiSSSE: System-wide security for searchable symmetric encryption. {\em Proceedings on Privacy Enhancing Technologies}. (2024).

\bibitem{guo2024gridse}
Guo, R., Li, J. \& Yu, S. GridSE: Towards practical secure geographic search via prefix symmetric searchable encryption (Full version). {\em arXiv preprint arXiv:2408.07916}. (2024).

\bibitem{hahn2018practical}
Hahn, F., Loza, N. \& Kerschbaum, F. Practical and secure substring search. {\em Proceedings of the 2018 International Conference on Management of Data}. pp. 163-176 (2018).

\bibitem{himeur2023ai}
Himeur, Y., Elnour, M., Fadli, F., Meskin, N., Petri, I., Rezgui, Y., Bensaali, F. \& Amira, A. AI-big data analytics for building automation and management systems: A survey, actual challenges and future perspectives. {\em Artificial Intelligence Review}. \textbf{56}, pp. 4929-5021 (2023).

\bibitem{hoang2023novel}
Hoang, C., Nguyen, M., Nguyen, T. \& Vu, H. A novel method for designing indexes to support efficient substring queries on encrypted databases. {\em Journal of King Saud University - Computer and Information Sciences}. \textbf{35}, pp. 20-36 (2023).

\bibitem{leontiadis2018storage}
Leontiadis, I. \& Li, M. Storage efficient substring searchable symmetric encryption. {\em Proceedings of the 6th International Workshop on Security in Cloud Computing}. pp. 3-13 (2018).

\bibitem{li2023survey}
Li, F., Ma, J., Miao, Y., Liu, X., Ning, J. \& Deng, R. A survey on searchable symmetric encryption. {\em ACM Computing Surveys}. \textbf{56}, pp. 1-42 (2023).

\bibitem{lambregts2022val}
Lambregts, S., Chen, H., Ning, J. \& Liang, K. Val: Volume and access pattern leakage-abuse attack with leaked documents. {\em European Symposium on Research in Computer Security}. pp. 653-676 (2022).

\bibitem{markatou2021reconstructing}
Markatou, E., Falzon, F., Tamassia, R. \& Schor, W. Reconstructing with less: Leakage abuse attacks in two dimensions. {\em Proceedings of the 2021 ACM SIGSAC Conference on Computer and Communications Security}. pp. 2243-2261 (2021).

\bibitem{mendonca2018data}
Mendonca, S. Data security in cloud using AES. {\em International Journal of Engineering Research and Technology}. \textbf{7} (2018).

\bibitem{naveed2015fallacy}
Naveed, M. The fallacy of composition of oblivious ram and searchable encryption. {\em Cryptology ePrint Archive}. (2015).

\bibitem{nick2007logistic}
Nick, T. \& Campbell, K. Logistic regression. {\em Topics in Biostatistics}. pp. 273-301 (2007).

\bibitem{nie2024query}
Nie, H., Wang, W., Xu, P., Zhang, X., Yang, L. \& Liang, K. Query recovery from easy to hard: Jigsaw attack against SSE. {\em 33rd USENIX Security Symposium (USENIX Security 24)}. pp. 2599-2616 (2024).

\bibitem{ning2021leap}
Ning, J., Huang, X., Poh, G., Yuan, J., Li, Y., Weng, J. \& Deng, R. LEAP: Leakage-abuse attack on efficiently deployable, efficiently searchable encryption with partially known dataset. {\em Proceedings of the 2021 ACM SIGSAC Conference on Computer and Communications Security}. pp. 2307-2320 (2021).

\bibitem{9165915}
Ning, J., Poh, G., Huang, X., Deng, R., Cao, S. \& Chang, E. Update recovery attacks on encrypted database within two updates using range queries leakage. {\em IEEE Transactions on Dependable and Secure Computing}. \textbf{19}, pp. 1164-1180 (2022).

\bibitem{8443434}
Ning, J., Xu, J., Liang, K., Zhang, F. \& Chang, E. Passive attacks against searchable encryption. {\em IEEE Transactions on Information Forensics and Security}. \textbf{14}, pp. 789-802 (2019).

\bibitem{noor2022simple}
Noor, S., Tajik, O. \& Golzar, J. Simple random sampling. {\em International Journal of Education \& Language Studies}. \textbf{1}, pp. 78-82 (2022).

\bibitem{oya2021hiding}
Oya, S. \& Kerschbaum, F. Hiding the access pattern is not enough: Exploiting search pattern leakage in searchable encryption. {\em 30th USENIX Security Symposium (USENIX Security 21)}. pp. 127-142 (2021).

\bibitem{10.1145/3064005}
Poh, G., Chin, J., Yau, W., Choo, K. \& Mohamad, M. Searchable symmetric encryption: Designs and challenges. {\em ACM Computing Surveys}. \textbf{50}(5) (2017). https://doi.org/10.1145/3064005

\bibitem{satapathy2016comprehensive}
Satapathy, A., Livingston, J. \& Others. A comprehensive survey on SSL/TLS and their vulnerabilities. {\em International Journal of Computer Applications}. \textbf{153}, pp. 31-38 (2016).

\bibitem{song2000practical}
Song, D., Wagner, D. \& Perrig, A. Practical techniques for searches on encrypted data. {\em Proceeding 2000 IEEE Symposium on Security and Privacy. S\&P 2000}. pp. 44-55 (2000).

\bibitem{strizhov2015substring}
Strizhov, M. \& Ray, I. Substring position search over encrypted cloud data using tree-based index. {\em 2015 IEEE International Conference on Cloud Engineering}. pp. 165-174 (2015).

\bibitem{xu2021interpreting}
Xu, L., Duan, H., Zhou, A., Yuan, X. \& Wang, C. Interpreting and mitigating leakage-abuse attacks in searchable symmetric encryption. {\em IEEE Transactions on Information Forensics and Security}. \textbf{16}, pp. 5310-5325 (2021).

\bibitem{xu2023leakage}
Xu, L., Zheng, L., Xu, C., Yuan, X. \& Wang, C. Leakage-abuse attacks against forward and backward private searchable symmetric encryption. {\em Proceedings of the 2023 ACM SIGSAC Conference on Computer and Communications Security}. pp. 3003-3017 (2023).

\bibitem{zhang2016all}
Zhang, Y., Katz, J. \& Papamanthou, C. All your queries are belong to us: The power of file-injection attacks on searchable encryption. {\em 25th USENIX Security Symposium (USENIX Security 16)}. pp. 707-720 (2016).

\end{thebibliography}
%
\newpage

\section{Appendix}\label{sec:appendix}
% \subsection{Algorithm 1: Unique Column-Sum Mapping}
\begin{algorithm}
\caption{Unique Column-Sum Mapping}
\label{alg:column_mapping_io}
\textbf{Input:} Partial $\mathbf{A'}$, full $\mathbf{B}$ \\
\textbf{Output:} Mappings $\mathcal{M}=\{(es_k, s_{y_k})\}$
\begin{algorithmic}[1]
\Function{ColumnSumMap}{$\mathbf{A'}, \mathbf{B}$}
    \State $\mathbf{A''} \gets \text{Extend}(\mathbf{A'}, m)$
    \State $S_B \gets \text{ColSums}(\mathbf{B})$, $S_A \gets \text{ColSums}(\mathbf{A''})$
    \For{$k \in [n]$ where $S_B[k]$ unique}
        \If{$\exists! k'$ with $S_B[k]=S_A[k']$}
            \State $\mathcal{M} \gets \mathcal{M} \cup \{(es_k, s_{y_k})\}$
        \EndIf
    \EndFor
    \State \Return $\mathcal{M}$
\EndFunction
\end{algorithmic}
\end{algorithm}

% \begin{algorithm}
% \caption{Occurrence Matrix Mapping}
% \label{alg:occurrence_mapping}
% \textbf{Input:} $\mathcal{M}$, $\mathbf{M}$, $\mathbf{M'}$, $\mathbf{A''}$, $\mathbf{B}$ \\
% \textbf{Output:} New mappings $S$
% \begin{algorithmic}[1]
% \Function{OccurrenceMap}{$\mathcal{M}, \mathbf{M}, \mathbf{M'}, \mathbf{A''}, \mathbf{B}$}
%     \Repeat
%         \State $S \gets \emptyset$
%         \For{each unmapped $s_{y_{j'}}$}
%             \State $c' \gets \text{Sum}(\mathbf{A''}[:,j'])$
%             \State $ED \gets \{es_j | \text{Sum}(\mathbf{B}[:,j])=c'\}$
%             \For{$(es_k,s_k) \in \mathcal{M}$}
%                 \State $ED \gets ED \setminus \{es_j | \mathbf{M}_{j,k}\neq\mathbf{M'}_{j',k}\}$
%             \EndFor
%             \If{$|ED|=1$} $S \gets S \cup \{(es_j,s_{y_{j'}})\}$ \EndIf
%         \EndFor
%         \State $\mathcal{M} \gets \mathcal{M} \cup S$
%     \Until{$S=\emptyset$}
%     \State \Return $S$
% \EndFunction
% \end{algorithmic}
% \end{algorithm}

\begin{algorithm}
\caption{Occurrence Matrix Mapping}
\label{alg:occurrence_mapping}
\textbf{Input:} $\mathcal{M}$, $\mathbf{M}$, $\mathbf{M'}$, $\mathbf{A''}$, $\mathbf{B}$ \\
\textbf{Output:} New mappings $S$
\begin{algorithmic}[1]
\Function{OccurrenceMap}{$\mathcal{M}, \mathbf{M}, \mathbf{M'}, \mathbf{A''}, \mathbf{B}$}
    \Repeat
        \State $S \gets \emptyset$
        \For{each unmapped $s_{y_{j'}}$}
            \State $ED \gets \{es_j | \text{Sum}(\mathbf{B}[:,j])=\text{Sum}(\mathbf{A''}[:,j'])\}$
            \For{$(es_k,s_k) \in \mathcal{M}$} \State $ED \gets ED \setminus \{es_j | \mathbf{M}_{j,k}\neq\mathbf{M'}_{j',k}\}$ \EndFor
            \If{$|ED|=1$} $S \gets S \cup \{(ED[0],s_{y_{j'}})\}$ \EndIf
        \EndFor
        \State $\mathcal{M} \gets \mathcal{M} \cup S$
    \Until{$S=\emptyset$} \State \Return $S$
\EndFunction
\end{algorithmic}
\end{algorithm}

% \begin{algorithm}
% \caption{Unique Row Mapping}
% \label{alg:row_mapping_io}
% \textbf{Input:} $\mathbf{B_c}$, $\mathbf{A''_c}$, $\mathcal{M}$ \\
% \textbf{Output:} Mappings $\mathcal{R}=\{(t_i,a_{i'})\}$
% \begin{algorithmic}[1]
% \Function{RowMap}{$\mathbf{B_c}, \mathbf{A''_c}, \mathcal{M}$}
%     \State $P_B \gets \text{RowPatterns}(\mathbf{B_c})$, $P_A \gets \text{RowPatterns}(\mathbf{A''_c})$
%     \For{each unique $p$ in $P_B$} \If{$\exists! i'$ with $P_A[i']=p$} \State $\mathcal{R} \gets \mathcal{R} \cup \{(t_i,a_{i'})\}$ \EndIf \EndFor
%     \State \Return $\mathcal{R}$
% \EndFunction
% \end{algorithmic}

\begin{algorithm}[t]
\caption{Unique Row Mapping}
\label{alg:row_mapping_io}
\textbf{Input:} $\mathbf{B_c}$, $\mathbf{A''_c}$, $\mathcal{M}$ \\
\textbf{Output:} Mappings $\mathcal{R}=\{(t_i,a_{i'})\}$
\begin{algorithmic}[1]
\Function{RowMap}{$\mathbf{B_c}, \mathbf{A''_c}, \mathcal{M}$}
    \State $P_B \gets \text{RowPatterns}(\mathbf{B_c})$, $P_A \gets \text{RowPatterns}(\mathbf{A''_c})$
    \For{each unique $p$ in $P_B$} \If{$\exists! i'$ with $P_A[i']=p$} \State $\mathcal{R} \gets \mathcal{R} \cup \{(t_i,a_{i'})\}$ \EndIf \EndFor
    \State \Return $\mathcal{R}$
\EndFunction
\end{algorithmic}
\end{algorithm}

\begin{algorithm}[t]
\caption{Unique Column Mapping}\label{alg:unique_column_mapping}
\begin{algorithmic}[1]
\Function{ColumnMap}{$\mathbf{B_r}, \mathbf{A''_r}, \mathcal{R}$}
\State $C_B,C_A \gets \text{ColPatterns}(\mathbf{B_r}),\text{ColPatterns}(\mathbf{A''_r})$
\For{$j \in \text{Unique}(C_B)$} \If{$|\{j'|C_A[j']=C_B[j]\}|=1$} $\mathcal{M}' \gets \mathcal{M}' \cup \{(es_j,s_{y_{j'}})\}$ \EndIf \EndFor
\State \Return $\mathcal{M}'$\EndFunction
\end{algorithmic}
\end{algorithm}
% \begin{algorithm}
% \caption{Unique Column Mapping}
% \label{alg:unique_column_mapping}
% \textbf{Input:} $\mathbf{B_r}$, $\mathbf{A''_r}$, $\mathcal{R}$ \\
% \textbf{Output:} New mappings $\mathcal{M}'$
% \begin{algorithmic}[1]
% \Function{ColumnMap}{$\mathbf{B_r}, \mathbf{A''_r}, \mathcal{R}$}
%     \State $C_B \gets \text{ColPatterns}(\mathbf{B_r})$, $C_A \gets \text{ColPatterns}(\mathbf{A''_r})$
%     \For{each unique $j$ in $C_B$} \If{$\exists! j'$ with $C_A[j']=C_B[j]$} \State $\mathcal{M}' \gets \mathcal{M}' \cup \{(es_j,s_{y_{j'}})\}$ \EndIf \EndFor
%     \State \Return $\mathcal{M}'$
% \EndFunction
% \end{algorithmic}
% \end{algorithm}

% \begin{algorithm}[t]
% \caption{Iterative Recovery}\label{alg:iterative_recovery}
% \begin{algorithmic}[1]
% \Function{IterativeRecover}{$\mathbf{B}, \mathbf{A''}, \mathcal{M}$}
% \State $\mathbf{B'},\mathbf{A'''} \gets \text{ZeroMatchedCols}(\mathbf{B},\mathcal{M}),\text{ZeroMatchedCols}(\mathbf{A''},\mathcal{M})$
% \Repeat \State $\mathbf{V_B},\mathbf{V_A} \gets \text{ColSums}(\mathbf{B'}),\text{ColSums}(\mathbf{A'''})$
% \For{$j \in \text{Unique}(\mathbf{V_B})$} \If{$|\{j'|\mathbf{V_A}[j']=\mathbf{V_B}[j]\}|=1$} $\mathcal{M} \gets \mathcal{M} \cup \{(es_j,s_{y_{j'}})\}$ \EndIf \EndFor
% \Until{$\neg\exists$ new matches} \Return $\mathcal{M}$\EndFunction
% \end{algorithmic}
% \end{algorithm}

\begin{algorithm}[t]
\caption{Iterative Recovery}
\label{alg:iterative_recovery}
\textbf{Input:} $\mathbf{B}$, $\mathbf{A''}$, $\mathcal{M}$ \\
\textbf{Output:} Augmented $\mathcal{M}'$
\begin{algorithmic}[1]
\Function{IterativeRecover}{$\mathbf{B}, \mathbf{A''}, \mathcal{M}$}
    \State $\mathbf{B'} \gets \text{ZeroMatchedCols}(\mathbf{B}, \mathcal{M})$, $\mathbf{A'''} \gets \text{ZeroMatchedCols}(\mathbf{A''}, \mathcal{M})$
    \Repeat
        \State $\mathbf{V_B} \gets \text{ColSums}(\mathbf{B'})$, $\mathbf{V_A} \gets \text{ColSums}(\mathbf{A'''})$
        \For{each unique $j$ in $\mathbf{V_B}$}
            \If{$\exists! j'$ with $\mathbf{V_B}[j]=\mathbf{V_A}[j']$}
                \State $\mathcal{M} \gets \mathcal{M} \cup \{(es_j,s_{y_{j'}})\}$ \text{ and zero columns $j$, $j'$}
            \EndIf
        \EndFor
    \Until{no new matches} \State \Return $\mathcal{M}$
\EndFunction
\end{algorithmic}
\end{algorithm}

\end{document}